\documentclass[apj]{emulateapj}
\usepackage{apjfonts}
\submitted{Received 2007 April 2; accepted 2007 July 15}
\journalinfo{THE ASTROPHYSICAL JOURNAL, Preprint}

\defcitealias{moore06}{MS06}

\begin{document}
\title{THE ERUPTION FROM A SIGMOIDAL SOLAR ACTIVE REGION ON 2005 MAY 13}
\author{CHANG LIU\altaffilmark{1,2}, JEONGWOO LEE\altaffilmark{2}, VASYL YURCHYSHYN\altaffilmark{1}, NA DENG\altaffilmark{2,3}, KYUNG-SUK CHO\altaffilmark{2,4}, MARIAN KARLICK{\'Y}\altaffilmark{5}, AND HAIMIN WANG\altaffilmark{1,2}}
\affil{1. Big Bear Solar Observatory, New Jersey Institute of Technology, 40386 North Shore Lane, Big Bear City, CA 92314-9672; cliu@bbso.njit.edu}
\affil{2. Center for Solar-Terrestrial Research, New Jersey Institute of Technology, University Heights, Newark, NJ 07102-1982}
\affil{3. Department of Physics and Astronomy, California State University, Northridge, CA 91330-8268}
\affil{4. Korea Astronomy and Space Science Institute, Daejeon 305-348, Korea}
\affil{5. Astronomical Institute of the Academy of Sciences of the Czech Republic, 25165 Ond{\v r}ejov, Czech Republic}

\begin{abstract}
This paper presents a multiwavelength study of the M8.0 flare and its associated fast halo CME that originated from a bipolar active region NOAA 10759 on 2005 May 13. The source active region has a conspicuous sigmoid structure at TRACE 171~\AA\ channel as well as in the SXI soft X-ray images, and we mainly concern ourselves with the detailed process of the sigmoid eruption as evidenced by the multiwavelength data ranging from H$\alpha$, WL, EUV/UV, radio, and hard X-rays (HXRs). The most important finding is that the flare brightening starts in the core of the active region {\it earlier} than that of the rising motion of the flux rope. This timing clearly addresses one of the main issues in the magnetic eruption onset of sigmoid, namely, whether the eruption is initiated by an internal tether-cutting to allow the flux rope to rise upward or a flux rope rises due to a loss of equilibrium to later induce tether cutting below it. Our high time cadence SXI and H$\alpha$ data shows that the first scenario is relevant to this eruption. As other major findings, we have the RHESSI HXR images showing a change of the HXR source from a confined footpoint structure to an elongated ribbon-like structure after the flare maximum, which we relate to the sigmoid-to-arcade evolution. Radio dynamic spectrum shows a type II precursor that occurred at the time of expansion of the sigmoid and a drifting pulsating structure in the flare rising phase in HXR. Finally type II and III bursts are seen at the time of maximum HXR emission, simultaneous with the maximum reconnection rate derived from the flare ribbon motion in UV. We interpret these various observed properties with the runaway tether-cutting model proposed by \citeauthor{moore01} in 2001.

\end{abstract}

\keywords{Sun: activity --- Sun: flares --- shock waves --- Sun: radio radiation --- Sun: UV radiation --- Sun: X-rays, gamma rays --- Sun: coronal mass ejections (CMEs)}

\section{INTRODUCTION}
Magnetic configurations that are favorable for eruption have been of recent interest in relation to space weather. One of the strongest candidates is the so-called sigmoid, an S-shaped magnetic field structure as seen in soft X-rays (SXRs). It was first investigated by \cite{rust96} who found that many large SXR brightenings associated with H$\alpha$ filament eruptions and coronal mass ejections (CMEs) had the sigmoidal shape. Several authors claimed that when active regions are in the sigmoid configuration, a higher probability of eruption to produce flares and associated CMEs is generally expected \citep{hudson98,canfield99,glover00}, and thus a sigmoid is an important precursor of a CME \citep{canfield00}. It was also found that a sigmoid often changes, after the eruption, to an arcade of loops, a process termed ``sigmoid-to-arcade'' evolution \citep{sterling00}. The sigmoid is now regarded as an important signature in space weather forecasts \citep{rust05}. Importance of sigmoid structure in coronal energy storage and release is also studied in terms of a flux rope model \citep{gibson06b}.

A qualitative model for sigmoid eruption was proposed by \citet{moore80} and further elaborated by \citet[][hereafter, Moore's model]{moore01}, in which a magnetic explosion is unleashed by internal tether-cutting reconnection between the highly sheared magnetic fields in the middle of the sigmoid. This first stage reconnection, which is the eruption onset, produces a low-lying shorter loop across the magnetic polarity inversion line (PIL) and a longer twisted loop connecting the two far ends of the sigmoid. The second stage begins when the formed twisted loop subsequently becomes unstable and erupts outward, distending the envelope field that overarches the sigmoid. The opened legs of the envelope field subsequently reconnect back to form an arcade structure and the ejecting plasmoid escapes as a CME. This model features a scenario of flare/CME initiating from internal reconnection deep in a bipolar active region, which is opposite to the breakout model \citep{antiochos98} involving multipolar connectivity and initial external reconnection in relation to remote brightenings \cite[see more discussions in][hereafter, MS06]{moore06}. Thus timing of initial flare brightening is useful in judging the most relevant eruption mechanism in the multipolar configuration \cite[e.g.,][]{yurchyshyn06a}. Although the Moore's model was proposed based on and mostly evidenced by the morphological change of flaring magnetic structure, support was also found by studying evolution of the photospheric magnetic fields around flaring PIL \citep{wang06}.

\begin{figure*}[t]
\epsscale{.8}
\plotone{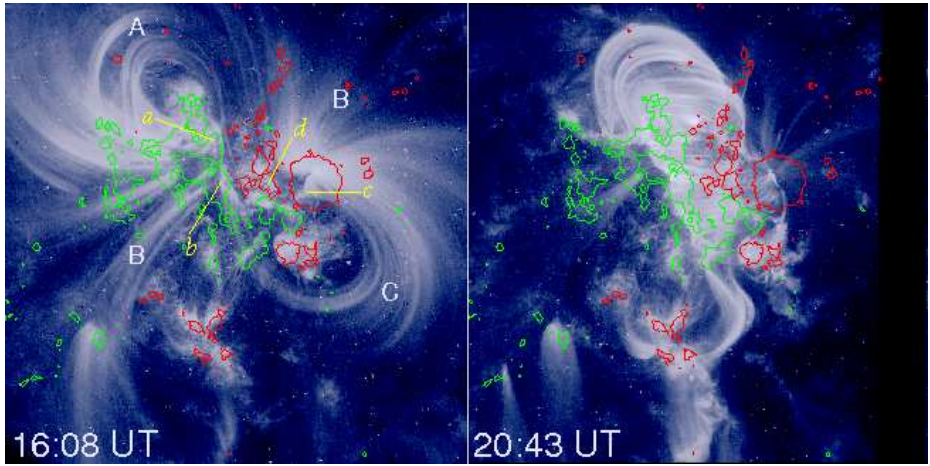}
\caption{Pre- and postflare images from TRACE 171~\AA\ channel showing the sigmoid-to-arcade evolution of the coronal magnetic field in the 2005 May 13 M8.0 event. ``A'' and ``C'' denote the magnetic elbows and ``B'', envelope loops, following the nomenclature used by \cite{moore01}. \textit{a-d} denote the footpoints of the sigmoidal fields. MDI longitudinal magnetic field is superimposed with the red and green contours representing positive and negative fields, respectively. The contour levels are $\pm$ 50~G. The field of view is 384\arcsec\ $\times$ 384\arcsec\ at N12, E11. \label{fig.1}}
\end{figure*}

To date, active region sigmoids have primarily been observed using the Yohkoh Soft X-ray Telescope \citep[SXT;][]{tsuneta91}, implying that they are at temperatures of 2~MK and higher. We, however, found that the active region NOAA 10759 (N12, E11) on 2005 May 13 appears in a sigmoid shape not only in SXRs but also in the EUV channel 171~\AA\ of the Transition Region and Coronal Explorer \citep[TRACE;][]{handy99}. The observation of sigmoid in EUV means that this structure can be seen at a wider range of temperatures down to 1~MK, and more importantly, the sigmoidal structure is much more clearly visible thanks to the higher spatial resolution (0.5\arcsec\ per pixel) of TRACE than that of SXT (2.45\arcsec\ per pixel in full-resolution mode). This sigmoid active region spawned a major Sun-to-Earth event with a flare classified as 2B/M8.0, an associated fast halo CME, and an intense geomagnetic storm on 2005 May 15. \citet{yurchyshyn06b} carried out a detailed study of this event to suggest how this sigmoid formed and how its eruption was related to the magnetic cloud observed near the Earth.

In this paper we present a comprehensive study of the 2005 May 13 event with focus on the eruption process of the sigmoid near the sun. In \S~2 we summarize the data sets used in this study. In \S~3, the main results of the multiwavelength data analysis are described. We determine the kinematics of the eruption in \S~4, and summarize the major findings in \S~5.

\section{OBSERVATIONAL DATA}
The 2005 May 13 flare started at 16:13~UT, reached its maximum at 16:57~UT, and ended at $\sim$17~UT on May 14 in GOES SXR flux, and thus is recorded as a long duration event. The event was well covered by many space- and ground-based instruments.

In addition to the TRACE 171~\AA\ channel, the active region EUV sigmoid is also obvious in the 195~\AA\ images obtained with the EUV Imaging Telescope \citep[EIT;][]{delaboudiniere95} on the Solar and Heliospheric Observatory \cite[SOHO;][]{domingo95} with 5.26\arcsec\ pixel resolution and $\sim$15~minutes cadence, representing an Fe\,{\sc xii} line formed at a temperature around 1.5~MK. Observation of the sigmoid in SXRs was made with the Solar X-ray Imager \citep[SXI;][]{hill05} on the GOES~12 satellite using the polyimide thin filter sensitive to the coronal temperature at 3.8~MK. We removed the instrument point-spread function from the images, which have a time cadence ranging from 1--4~minutes and a pixel resolution of 5\arcsec.

The TRACE 1600~\AA\ channel covered this event with a highest cadence of 3~s during some intervals and 0.5\arcsec\ pixel resolution, which is used to study the flare ribbon motion in this event. Flare emission observed in the 1600~\AA\ band comes predominantly from the upper chromosphere and transition region and is thought to be produced by a mixture of particle precipitation and thermal conduction \citep[see, e.g.,][]{warren01}. The photospheric magnetic field of the active region was measured with the Michelson Doppler Imager \cite[MDI;][]{scherrer95} on SOHO. We also obtained hard X-ray (HXR) lightcurves and images from the Reuven Ramaty High Energy Solar Spectroscopic Imager \cite[RHESSI;][]{lin02} to explore the high energy release in this event.

Full-disk and high-resolution H$\alpha$ images were obtained at the Big Bear Solar Observatory (BBSO) by the 20~cm and 25~cm refractors, respectively. We used H$\alpha$ images with pixel resolution of $\sim$0.6\arcsec\ to monitor the evolution of the active region filament during this event.

Radio observations of this event were made at the Owens Valley Solar Array (OVSA), the Ond\v{r}ejov radiospectrograph \citep{jiricka93}, the Tremsdorf Solar Radio Observatory of the Astrophysical Institute (AI) Potsdam \citep{mann92}, and the Green Bank Solar Radio Burst Spectrometer \citep[GBSRBS;][]{white06}. All these data are digitally recorded. The Ond{\v r}ejov data covered the whole event at 0.8--2.0~GHz and only the early phase at 2.0--4.5~GHz with 0.1~s time resolution. The AI Potsdam instrument consists of swept-frequency spectrographs in the ranges 40--90, 100--170, 200--400, and 400--800 MHz, with a sweep rate of 10~s$^{-1}$. There are, however, data gaps in 100--170 and 400--800~MHz ranges during this event. The GBSRBS data used in this study were obtained with its low frequency system composed of a stand-alone active dipole that operates at approximately 20--70~MHz with 1~s sampling cadence.

A full-halo CME associated with this event was detected by the Large Angle and Spectrometric Coronagraph \cite[LASCO;][]{brueckner95} on SOHO. Due to the failure of the LASCO/EIT Electronics Box, only one C2 and one C3 full-frame images were recorded to show the halo CME. By combining three other C3 partial frames, we managed to follow the CME development in the southwest direction. 

\begin{figure*}[t]
\epsscale{0.8}
\plotone{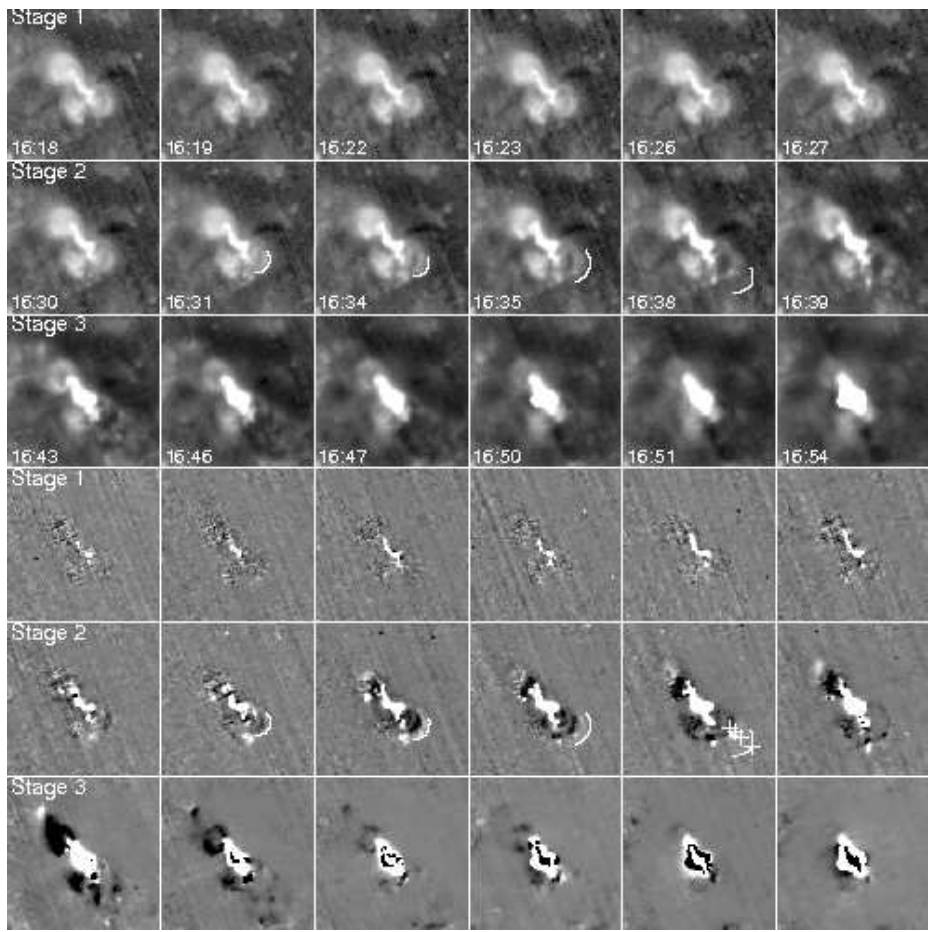}
\caption{Time sequence of SXI SXR images (upper three rows) showing the evolution of the sigmoid that exhibits three stages (see discussion in \S~3.1.2). Lower three rows are the running difference images. The outermost fronts of the southwest elbow are outlined to show its expansion, with crosses denoting the measured positions at $\sim$238$^{\circ}$. The field of view is 700\arcsec\ $\times$ 700\arcsec. West is to the right and solar north is up. \label{fig.2}}
\end{figure*}

\section{RESULTS AND ANALYSIS}
In this section, we describe major observational features found in all data sets. In specific, we present the sigmoid structure and its evolution observed at EUV and SXRs, chromospheric filament, CME and associated coronal dimmings in WL and EUV, respectively, eruption signatures in radio dynamic spectra, and finally ribbon motions in UV and HXRs. We also interpret the results and discuss their implication under the context of the Moore's model whenever a plausible comparison can be made.

\subsection{Morphological Evolution of Sigmoid}

\subsubsection{Sigmoid in EUV Images}
We show, in Figure~\ref{fig.1}, TRACE 171~\AA\ images taken just before the event (16:08~UT, left panel) and in the postflare state (20:43~UT, right panel). The superposed contours are longitudinal fields measured with the MDI magnetogram, which show that this is a bipolar active region consisting of a main round sunspot at the leading side with positive magnetic polarity and a trailing part of negative polarity. The preflare TRACE EUV image exhibits a very symmetric sigmoid rooted at four footpoints denoted as \textit{a}--\textit{d} without ambiguity, unlike the much less obvious SXR sigmoids in many cases \citep[see examples in][]{sterling00}. Following the nomenclature in the Moore's model, we denote the two oppositely curved magnetic elbows as ``A'' and ``C'', each of which links one polarity to the other \citep[for more details on the magnetic configuration see][]{yurchyshyn06b}. They loop out on opposite ends of the PIL to form a typical sigmoid. In the middle of the sigmoid, its legs are clearly seen to be highly sheared along the PIL. The envelope field (denoted as ``B'') is much less sheared and extends outward, possibly overarching and tying down the sheared core field. After the eruption, the sigmoidal field changed to loops of an arcade, thus exhibiting the sigmoid-to-arcade evolution. Since the active region is located very close to the disk center, the images in this figure serve as a top view of the pre- and postflare configurations, and they coincide with those depicted in the Moore's model \cite[see Fig.~1 of][]{moore01}. In TRACE images, this sigmoid shape could be seen even four hours before the flare/CME \citep[c.f. Fig.~6 of][]{yurchyshyn06b} and the two magnetic elbows became more obvious as nearing the flare time. A similar evolution of the sigmoid can also be seen in the EIT 195~\AA\ images, at a lower spatial resolution.

\subsubsection{Sigmoid in SXR Images} \label{sxr}
We re-check the sigmoid structure with the SXR images from SXI that cover the flare impulsive phase missed by TRACE. In view of \citetalias{moore06}, it is necessary to investigate the eruption process with a sufficiently high cadence, because the relative timing of initial flare brightening to flux rope motion is critical to identify of the triggering mechanism. We take advantage of the favorable location of the event and a good cadence of SXI images to trace the sigmoid eruption process in detail. Figure~\ref{fig.2} shows the time sequence of SXI images across the flaring interval in the upper three rows and their running difference images in the lower three rows. We describe the sigmoid evolution in three distinct stages as follows.

Stage 1: From the first row of Figure~\ref{fig.2} ($\sim$16:18--16:27~UT), we can see that the flare core get gradually brighter at beginning of the event, which is more clearly visible in the difference images. We identify this brightening with the onset of the core reconnection, which indicates that the eruption begins with the internal reconnection between sheared fields in the middle of the sigmoid. This point is corroborated by the chromospheric flare ribbon emission discussed in \S~3.4.1 (also c.f. Fig.\ref{fig.25}).

Stage 2: The images in the second row show the loop expansion/rising phase ($\sim$16:30--16:39~UT). Two magnetic elbows lying northeast and southwest of the active region begin to expand outward, which is again more clearly visible in the difference images. The southwestern elbow evolves obviously with the outermost fronts outlined. The expansion motion of sigmoid elbows was also observed in the events studied by \cite{moore01} and interpreted as the beginning of the ejective eruption.

Stage 3: The third row shows the explosion phase (after $\sim$16:39~UT), in which the large-scale loops in both northeast and southwest direction were abruptly ejected outward, simultaneous with the sudden enhancement of the flare core emission. This is also evident from the difference images. In this stage, the envelope fields of the bipole appeared to be blown by the twisted flux rope as a result of the reconnection among the sigmoidal field. In the end they gradually close back to form a long-duration bright arcade.

A noteworthy finding here is the flare core brightening in stage 1. This implies that the magnetic reconnection in the sigmoid core occurred well before the rising motion of the flux rope in stage 2. This timing rules out the hypothesis that the flux rope rises due to a loss of equilibrium and then causes a magnetic reconnection behind. We also note that no remote brightenings are seen around this isolated simple bipolar region and therefore the external tether-cutting reconnection mechanism, i.e., the magnetic break-out model can also be rejected \citepalias[see related discussions in][]{moore06}. The above observation can be explained by the internal tether-cutting model, in which the magnetic reconnection starts first in the sigmoid core to cut out the tethers (field lines tied to the photosphere) by the amount enough to allow the flux rope to rise upward. Then the magnetic pressure of the expanding flux rope works as the driver of the eruption \citepalias{moore06}.

As to how the reconnection between the sheared sigmoid legs in the core field start in the first place, \citetalias{moore06} proposes that photospheric flows can slowly push them together to form the current sheet needed for reconnection onset. This kind of long-term converging flows are indeed detected for this event by \citet{yurchyshyn06b}. Detailed study of the reconnection initiation is however out of the scope of this paper.

\subsection{Filament, CME, and Coronal Dimming}
\begin{figure}[t]
\epsscale{1.15}
\plotone{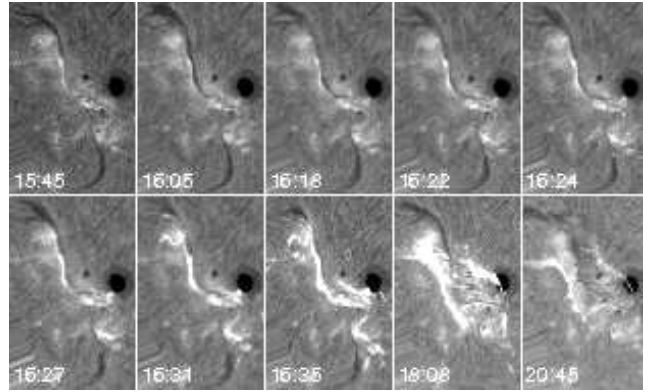}
\caption{BBSO high-resolution H$\alpha$ images co-temporal with the first two stages of the evolution in SXR (see Fig.\ref{fig.2}) as well as in the preflare and postflare status, showing the overall undisrupted filament between flare ribbons.} \label{fig.25}
\end{figure}

We show in Figure~\ref{fig.25} a sequence of H$\alpha$ images co-temporal with the first two stages of the evolution in SXR as well as in the preflare and postflare stages. There was an extended inverse S-shaped chromospheric filament running along the PIL, which is associated with the preflare EUV/SXR sigmoid. Prior to the eruption, it is noticeable that the filament became darker and bigger (c.f. images at 15:45 and 16:05~UT). During the early development of the H$\alpha$ ribbons ($\sim$16:18--16:35~UT), the filament appeared mostly undisturbed. This is best seen in 16:27~UT when the flare ribbons already formed while the filament remained almost the same as in 16:05~UT. Afterwards, during the flare impulsive phase, the northern part of the filament moved upward then fell back, and other parts briefly disappeared from the field of view possibly due to flare heating. Later on, most parts of the filament re-appeared in the postflare stage (see images at 18:08 and 20:45~UT). According to H$\alpha$ data, we hence conclude that the filament between the flare ribbons was overall undisrupted or may erupt only partially. We also note another filament in the southern field of view of Figure~\ref{fig.25} that erupted after the flare peak. However, it is not associated with the flare PIL and its eruption was most likely resulted from the flare disturbance.

In fact, it has been found that although the sigmoid-to-arcade transformation indicates a catastrophic perturbation in the corona, the active region filament lying below the sigmoid often shows no significant changes throughout the flare/CME event \citep{pevtsov02} and may erupt only partially \citep[][and references therein]{gibson06a}. A simple explanation is that the low lying loops resulting from the sigmoid core reconnection protect the filament from a disruption \citep{pevtsov02} unlike the classical flare models \citep[e.g.,][]{kopp76}, in which erupting filament serves as the trigger. The presence of non-erupting filaments in the core of this active region supplies another evidence for the internal tether-cutting reconnection beginning early in the eruption.

\begin{figure*}[t]
\epsscale{0.8}
\plotone{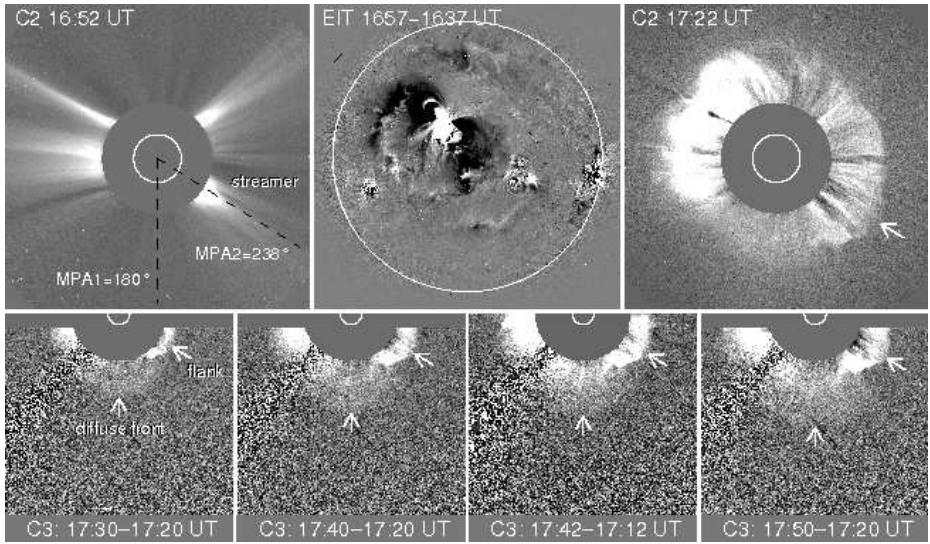}
\caption{LASCO images showing the evolution of the halo CME associated with the 2005 May 13 M8.0 flare. The preflare C2 image at 16:52~UT shows several streamer structure with the strongest one lying in the southwest. 17:22~UT C2 and 17:42~UT C3 are full-frame images, while C3 images at 17:30, 17:40, and 17:50~UT are partial-frame. The evolution of the outermost part of the CME diffuse front (CME leading edge) and the CME flank (where the CME interacts with the southwestern streamer and the intensity gradient is maximum) are both measured (marked by arrows), at measured position angles (MPA) of $\sim$180$^{\circ}$ and $\sim$238$^{\circ}$, respectively. The {\it top middle} panel is the SOHO EIT 195~\AA\ difference image showing the twin dimmings after the launch of the CME.} \label{fig.3}
\end{figure*}

As there is no noticeable filament eruption, it is appropriate to infer that the CME associated with this event represents a flux rope system resulting from the flare core reconnection as in Moore's model. Figure~\ref{fig.3} shows the development of the detected full-halo CME, which was first seen in C2 at 17:22~UT already half way across the C2 field of view and completely surrounded the C2 occulter. The C3 image taken at 17:42~UT also shows a very symmetric and bright halo \citep[see Fig.~8 of][]{yurchyshyn06b}. 
There are three C3 partial frames at 17:30, 17:40, and 17:50~UT, which we also used to measure the CME propagation speed. The outermost part of the CME diffuse front (CME leading edge) was measured at the position angle of $\sim$180$^{\circ}$, and the maximum intensity gradient was found at a position angle of $\sim$238$^{\circ}$, where the CME flank interacted with the southwestern streamer (see the preflare C2 image at 16:52~UT). The traced positions are pointed to with \textit{arrows}, from which we found the average speed of the CME leading edge and flank to be $\sim$1600~km~s$^{-1}$ and $\sim$1100~km~s$^{-1}$, respectively. Similar diffuse arc in front of the main CME was also reported before \cite[e.g.,][]{gary82}. Note that the LASCO Web site reports the speed of this CME to be $\sim$1689~km~s$^{-1}$ at the position angle of 2$^{\circ}$, which is very similar to our result for the southern diffuse front and thus might be the speed of the northern diffuse front that can also be seen in the full frame LASCO images. We remark, however, that the LASCO Web site only measures two frames and one of them, the C3 image at 17:12~UT, barely shows the CME.

We looked for a coronal dimming, an important signature of CMEs \cite[see, e.g.,][]{thompson00}, using EIT 195~\AA\ images. The difference image between just before and after the flare impulsive phase at 16:37 and 16:57~UT, respectively, is shown in the middle upper panel in Figure~\ref{fig.3}. We found two regions of strong dimmings that lie on opposite ends of the PIL, extending toward northeast and southwest. These dimmings can be naturally explained by the depletion of coronal material due to eruption of the sigmoidal fields following the Moore's model. This kind of so-called twin dimmings has been only occasionally observed in EIT \citep[see, e.g.,][]{thompson98}. In this case, the twin dimmings were not transient, but sustained after the eruption, as we can tell by comparing all the other postflare EIT images (17:07, 17:27, and 17:37~UT) with the preflare image. By 22:57~UT (EIT 195~\AA\ data has a gap between 17:37 and 22:57~UT), the southwestern dimming region already developed into an elongated trans-equatorial coronal hole.

\subsection{Radio Signatures}
\subsubsection{Type II and III Bursts}
\begin{figure}
\epsscale{1.2}
\plotone{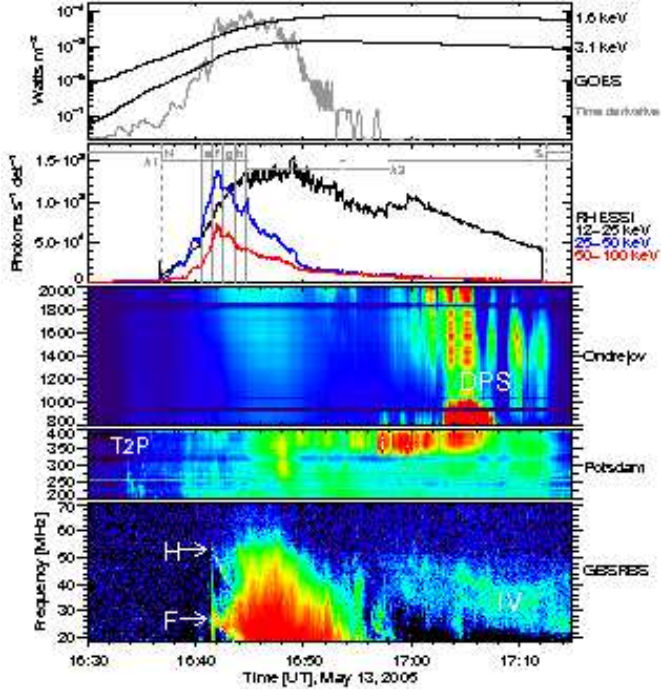}
\caption{Top panel: Time evolution of the GOES X-ray fluxes and the time derivative of the 1.6~keV band. Second panel: RHESSI photon rates binned into 4~s intervals. The 25--50 and 50--100~keV rates are times 20 and 60, respectively, for a clearer representation. The time intervals $e$--$h$ divided by the vertical lines are for RHESSI images shown in Fig.~\ref{fig.7}. The attenuator status for RHESSI switched between A1 and A3 during the observation period. ``N'' and ``S'' denote the time period of RHESSI night and south Atlantic anomaly, respectively. Third to bottom panels: Radio spectrum observed by Ond{\v r}ejov, Potsdam, and GBSRBS, respectively. Note the broadband DPS at 16:58--17:12~UT in the 1.0--2.0~GHz range, the concurrent decametric type IV burst in the 20--50~MHz range, and the metric type II precursor (``T2P'') around 16:34~UT in the 200--300~MHz range. Fundamental (``F'') and harmonic (``H'') components of the type II emissions detected by GBSRBS are depicted by the dotted lines. \label{fig.4}}
\end{figure}

It has been well known that type II radio bursts are the manifestation of shock waves in the solar corona, usually associated with either large flares or fast CMEs \citep{nelson85}. We thus studied type II burst in this event to infer the evolution of the ejecting flux rope. We show, in Figure~\ref{fig.4}, the GBSRBS radio spectra along with the X-ray light curves for this event, in which a fast decametric type II burst is clearly seen. We mark the fundamental and harmonic emissions with dotted lines in the bottom panel. The type II burst began at $\sim$16:41:30~UT with starting frequency $\sim$50~MHz (harmonic lane) and was preceded by a type III radio burst, which is conventionally interpreted as accelerated electrons escaping along open field lines. The occurrence of the type II and type III radio bursts are almost simultaneous with the peak of RHESSI 25--100~keV HXR emissions at $\sim$16:42~UT (second panel). The harmonic lane of the type II burst and the type III burst were also recorded by the dynamic spectrum from Potsdam in the 40--90~MHz band. In order to determine the speed of type II radio burst, we need to convert the frequency drift seen in the spectrograph to the trajectory of the type II source using a coronal electron density model. It has been suggested that the one-fold Newkirk model \citep{newkirk61} well represents the density in the inner corona and that the Mann model \citep{mann99} is better at coronal heights greater than $1.8~R_s$ \citep{warmuth05}. We followed this suggestion to estimate the speed of the type II radio burst to be $\sim$1200~km~s$^{-1}$, and the formation height of the metric type II precursor (discussed next in \S~3.3.2) in this event to be $\sim$0.9--1.7~$\times$~10$^{5}$~km (see Figure~\ref{fig.8}).

\subsubsection{Type II Precursors}
A remarkable feature at the very beginning of this flare is the radio bursts in the $\sim$200--300~MHz frequency range at 16:33:30--16:34:30~UT as detected in the Potsdam radiospectrogram (Fig.~\ref{fig.4}, the fourth panel). These fast drift bursts or pulsations are a type of radio emission called type II precursor \citep{klassen99}. In this event they are concurrent with the earliest energy release at $\sim$16:32--16:35~UT (traced here by the temporal derivative of the SXR light curve drawn as gray curve in top row of Fig.~\ref{fig.4}) and the onset of the SXR loop expansion (see \S~3.1.2). We suggest in \S~4 based on the similar timing that the observed type II precursor might be the radio signature of the ejective eruption manifested by the expansion/rising of the sigmoid elbows.

\begin{figure}[t]
\epsscale{1.15}
\plotone{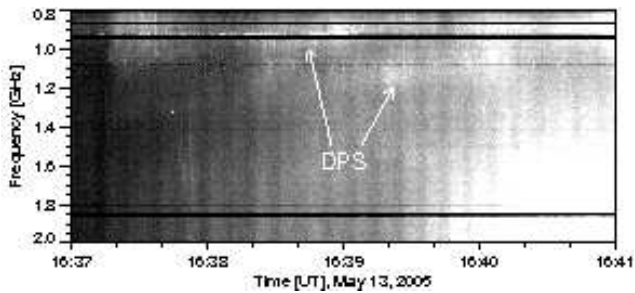}
\caption{A detailed view of the Ond{\v r}ejov radiospectrograph showing the narrowband DPSs in 0.8--1.4~GHz at 16:37:15--16:41:00~UT. \label{fig.5}}
\end{figure}

\subsubsection{Drifting Pulsating Structures}
At 16:37:15--16:41:00~UT in the 0.8--1.4 GHz range two weak narrowband drifting pulsating structures (DPS) were observed by the Ond{\v r}ejov spectrograph (see Fig.~\ref{fig.5}). Their frequency drift rate was about $\sim-$2.5~MHz~s$^{-1}$. These DPSs were observed during the rapid rising phase of the HXR emission, therefore they may represent the radio emission from ejected plasmoids formed during the flare reconnection process as proposed by \cite{kliem00} and \cite{karlicky05}.

In the postflare phase (16:58--17:12~UT), the 1.0--2.0~GHz range broadband DPS was observed by Ond{\v r}ejov (see Fig.~\ref{fig.4}, third panel) and OVSA, and can be associated with the postflare growing loop system \citep{svestka87}. In this event, the emission frequency of postflare DPS drifts at a rate $\sim-$1.2~MHz~s$^{-1}$ and it consists of several strong pulses with the characteristic period of about 90~s, which might indicate an oscillation of the postflare loops. Namely, this radio emission is probably generated by plasma emission processes at the top of the growing postflare loop system while there is still a slow reconnection above the postflare loops to inject the radio emitting particles \citep[see e.g.,][]{akimov96}. The radio drift toward lower frequencies is then explicable in terms of the decreasing electron density at the top of this loop with time. The termination shock \citep{aurass04} may also be present at the top of this loop system and contribute to the radio emission. We also note the concurrent decametric type IV burst as seen in GBSRBS data, and we speculate its source to be located well above the top of the postflare arcades, possibly in the upper part of the reconnecting current sheet.

\subsection{Dynamics of Chromospheric Flare Emissions}
\begin{figure}[t]
\epsscale{1.15}
\plotone{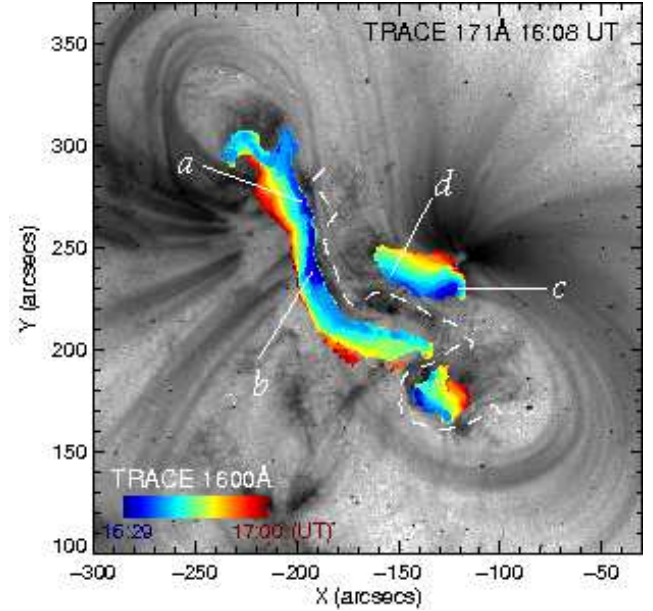}
\caption{Separation of flare ribbons from 16:29 to 17:00~UT observed in TRACE 1600~\AA\ (color-coded areas). The gray background is a preflare TRACE 171~\AA\ image at 16:08~UT showing the sigmoidal fields. The white dashed line is the PIL. Same marks of \textit{a}--\textit{d} as in Fig.~\ref{fig.1} are used to denote the initial four flare ribbons. \label{fig.6}}
\end{figure}

\subsubsection{UV Ribbon Motion}
Recently many studies were made to use flare ribbon separation motion as a measure for the magnetic reconnection in corona \citep[][and references therein]{saba06}. It is also shown that the magnetic reconnection rate derived in this way timely correlates with the flux rope evolution \cite[e.g.,][]{qiu04}. For the purpose of our sigmoid study, \citetalias{moore06} depicts the Moore's model by dividing the eruption process into two phases of ``slow runaway reconnection'' and ``explosive reconnection'' based on how fast the reconnection is, although there is no necessary different physics between them. Thus there is a need to quantitatively characterize the reconnection process.

We use the approach mentioned above, in which the change rate of the magnetic flux in the chromospheric ribbon area, $R=d\Phi/dt$, is measured and then regarded as equivalent to the rate at which magnetic flux is brought into the magnetic reconnection region in the corona. In this event the flare ribbon motion is clearly seen in TRACE 1600~\AA\ images, and we measure the ribbon area in every 15~s while the original TRACE data have 3~s cadence with some time gaps. We registered the successive flare ribbon position onto the co-aligned magnetogram and followed the intensity-based binary masks method presented by \cite{saba06} to determine $R$. The intensity threshold was set to a value of 5000 for easy detection of ribbon separation motion. This specific threshold value does not affect the overall timing (e.g., the peak time) of the derived $R$.

We depict the separation motion of the flare ribbons from $\sim$16:29--17:00~UT as color-coded areas in Figure~\ref{fig.6} (note the earliest flare brightenings in TRACE 1600~\AA\ can be seen at $\sim$16:20~UT). The flare brightenings are clearly seen lying at the foot of the sigmoidal fields as four flare ribbons marked as \textit{a}--\textit{d} (same marks as in Fig.~\ref{fig.1}; also c.f. Fig.~\ref{fig.25}). Later on, two main separating ribbons can be seen away from the PIL. This progression from four to two chromospheric ribbons vigorously implies that the eruption was initiated by the internal reconnection between the sigmoid elbows as elaborated in \citetalias{moore06}.

We plot the resulting $R$ in both the positive and negative magnetic fields as a function of time in Figure~\ref{fig.8} (middle panel). In our result the flux change in two polarities are well balanced and $R$ reaches its maximum at $\sim$5.1~$\times$~10$^{18}$~Mx~s$^{-1}$. We therefore can infer that the flux rope expansion/rising phase between $\sim$16:29--16:39~UT (see in \S~3.1.2) correspond to the ``slow runaway reconnection'' with increasing $R$, while the ``explosive reconnection'' occurs near the flare peak with the maximum $R$.

After time integration of $R$, the total amount of magnetic flux participating in the reconnection is $\sim$3.4~$\times$~10$^{21}$~Mx. We note that this value is about two times smaller than that reported by \citet{qiu05}, in which 1 minute cadence H$\alpha$ images were used. We attribute the difference mainly to the higher cadence TRACE UV data used in the present study and the different threshold value set to define the newly brightened flare patches. We also found that the magnetic flux reconnection rate is temporally correlated with both the HXR light curves and the time derivative of the SXR light curves, the signatures of nonthermal flare emissions. This coincidence is theoretically expected under the standard magnetic reconnection in bipolar magnetic structure \citep{priest02}.

It is, however, worth mentioning that the flare ribbons in the positive magnetic polarity actually broke into two parts, with the shorter one along the kink seen in the southwest end of the PIL, which was included in the early brightening in SXR in the sheared core filed (see Fig.\ref{fig.2}). Inspection of MDI magnetogram shows that this PIL kink is a persistent feature existed long before the flare. There was also no rapidly emerging magnetic flux reported to be associated with this event \citep{yurchyshyn06b}. Hence it is unlikely that this magnetic kink is related to the initiation of the eruption. More complicated three-dimensional magnetic field structure must be involved in producing flare emissions in this kinked region, which we did not incorporate into the tether-cutting scenario of this event (see Fig.\ref{fig.9}).

\begin{figure}[t]
\epsscale{1.15}
\plotone{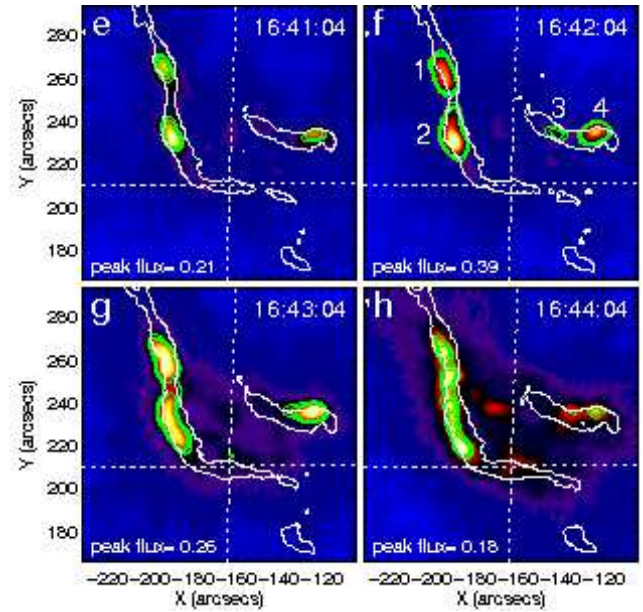}
\caption{A time sequence of RHESSI 25--100~keV HXR images integrated in the one-minute time intervals $e$--$h$ (denoted in Fig.~\ref{fig.4} second panel). Each RHESSI image was reconstructed with the CLEAN algorithm using grids 1--9 with the natural weighting scheme (giving $\sim$5.9\arcsec\ FWHM resolution). The peak flux in each image is labeled and the green contours show flux at levels of 0.11, 0.13, and 0.15 photons~cm$^{-2}$~s$^{-1}$~arcsec$^{-2}$. The white contours outline the TRACE 1600~\AA\ ribbons taken near the center of each RHESSI time interval. \label{fig.7}}
\end{figure}

\subsubsection{HXR Emissions}
Study the high energy release during flare process is another important approach to probe the reconnection process. RHESSI has an almost complete coverage of the impulsive phase of this event. We chose four one-minute time intervals ($e$--$h$ as denoted in Fig.~\ref{fig.4}, second panel) and reconstructed RHESSI images with the CLEAN algorithm using grids 1--9 with the natural weighting scheme, which gives $\sim$5.9\arcsec\ FWHM resolution. Figure~\ref{fig.7} shows the RHESSI 25--100~keV maps superposed with contours at fixed photon flux levels and those outlining TRACE UV ribbons. In the rise and maximum phase (intervals $e$--$f$), the HXR emissions appear as point-like compact sources located within the flare ribbons. After the flare maximum (intervals $g$--$h$), the HXR sources, however, become elongated to form a ribbon structure. The ribbon-like HXR sources are rarely reported in the past except \cite{masuda01}. The aforementioned footpoint-to-ribbon evolution of HXR emissions is more evident for the much stronger eastern HXR sources.

We found four HXR kernels in the flare maximum time with field strengths about 2--3 times larger than that of the other part of the ribbons without HXR emissions. This result agrees to the earlier result by \cite{asai02} who interpreted the confined HXR kernels as due to more intense energy release into the stronger fields of flare ribbons. However, the elongated HXR ribbon sources seen in the time interval $g$ and $h$ do not share this property \citep[see][for more details]{liu07,jing07}.

To understand why the ribbon-like HXR source appears in this event, we propose the following scenario based on the Moore's model. Until the flare maximum, the tether-cutting reconnection occurs between the two elbows in the middle of the sigmoid, during which the HXR sources can appear footpoint-like as usual. Although only until flare maximum can we resolve four HXR sources that lie at the footpoints of the sigmoidal loops, HXR emission should occur at the four footpoints of these loops at the beginning of the reconnection when the initial four H$\alpha$ ribbons appear as discussed in \S~3.4.1. This has to do with the fact that the initial flare brightening sites have weaker magnetic field meaning less HXR production, and RHESSI can only generate image when sufficient counts have been achieved. After the flare maximum, however, the magnetic envelope is blown out and the opened legs of the envelope will continue to reconnect back to form an arcade structure. The electrons accelerated in the magnetic arcade can bombard the chromosphere along the footpoints of the arcade, resulting in the ribbon-like HXR emissions. We therefore conclude that this seamlessly four HXR footpoints to two ribbon-like HXR sources transformation of the HXR source morphology can be another forceful indication of the sigmoid-to-arcade evolution of the magnetic field configuration.

\begin{figure}[t]
\epsscale{1.15}
\plotone{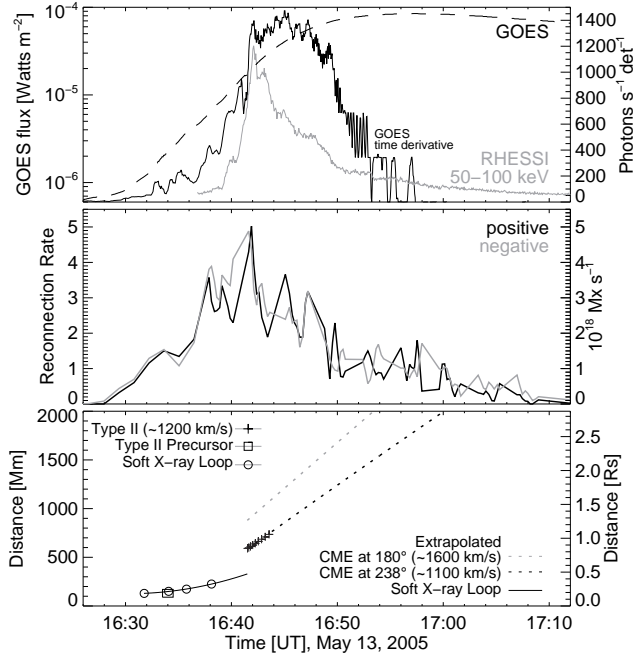}
\caption{Top panel: Flare light curves of GOES 1--8~\AA\ channel, its time derivative (black line), and RHESSI 50--100~keV photon rates binned into 4~s intervals. Middle panel: Magnetic reconnection rate derived in positive and negative magnetic fields. Bottom panel: Time evolutions of the distances of the expanding SXR loop, the type II radio burst, and the CME. The vertical axis represents the projected distances from the center of the sigmoid for the expanding SXR loop, the distances from the photosphere calculated with the Mann's and Newkirk's coronal electron density model for the type II burst and type II precursor, respectively, and the heights of the CME front and flank by extrapolating a constant velocity and constant deceleration best-fit, respectively, back to the flare site. The size of the symbol for the metric type II precursor denotes its range of appearance time and formation height. Note that the CME leading edge was at $\sim 5.2~R_\sun$ at the time of the first LASCO C2 image at 17:22~UT. ``Rs'' represents the solar radius. \label{fig.8}}
\end{figure}

\section{Magnetic Eruption}
We are now ready to quantitatively determine the rising motion of the magnetic flux rope, although this is a disk event and the height of the flux rope is not directly measured. We can instead use the horizontal motion of the expanding elbows as a proxy for the height of the ascending flux rope assuming a spherical symmetry. The projected distance of the SXR loop front is measured from the expanding southwestern elbow depicted in Figure~\ref{fig.2} ({\it crosses}) and the center of the sigmoid and is fitted assuming a constant acceleration. The heights of the decametric type II burst and the metric type II precursor are estimated using Mann's and Newkirk's coronal electron density models, respectively. The best-fits of height-time data of the southern CME leading edge (at 180$^{\circ}$) and flank (at 238$^{\circ}$) were extrapolated back to the flare site assuming a constant velocity and a constant deceleration, respectively. The heights of the expanding SXR loop, the type II radio burst and precursor, and the CME determined in these ways are shown in Figure~\ref{fig.8} (bottom panel), in comparison with the time profile of the magnetic reconnection (middle panel), the light curves of GOES SXR flux at the 1--8~\AA\ channel, its time derivative, and RHESSI 50-100~keV photon rates (top panel). Based on these results, we finally organize the observational evidence for magnetic eruption to determine its kinematics as follows.

First, the radio and soft-X ray data show that the formation height of the metric type II precursor is similar to the height of the cotemporal rising flux rope (see Fig.~\ref{fig.8} bottom panel). It is thus likely that this type II precursor is driven by the moving X-ray loops, and therefore is a signature for the onset of shock formation in the low corona \cite[see][and references therein]{dauphin06}. We however remark that the X-ray loop in this event moved at $\sim$250~km~s$^{-1}$, which might not be high enough to generate the shock in the low corona ($\sim$10$^5$~km). But the velocity might have been underestimated since we could only measure the projected distances. As a comparison, type II emissions are usually associated with rapidly rising X-ray structure \cite[e.g.,][]{gopalswamy97,klein99,dauphin06}.

Second, the decametric type II burst shows a similar height-time evolution to that of the southwestern CME flank, where it interacts with a strong streamer structure. The dense helmet streamer compared to the location of the southern diffusion front forms a low Alfvenic region as favorable for the generation of the type II burst. This therefore suggests that the type II radio emission in this event was originated when the CME interacted with the dense coronal streamer, as has been previously reported \citep[e.g.,][]{vanderholst02,reiner03,mancuso04,cho07}.

\begin{figure}[t]
\epsscale{0.9}
\plotone{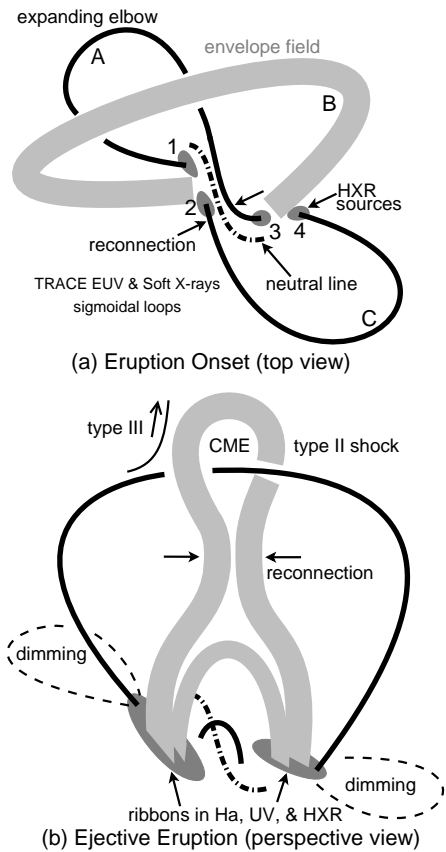}
\caption{Schematic picture interpreting our multiwavelength observations, based on the eruptive model for sigmoidal bipoles proposed by \cite{moore80} and elaborated in \cite{moore01}. See detailed discussions in \S~5. \label{fig.9}}
\end{figure}

Third, at the flare peak ($\sim$16:42~UT), we see the maximum of the magnetic reconnection rate and the sudden increase of height of flux rope, together. We are not sure whether there was a true acceleration of the flux rope, due to the data gap around this time. As a comparison, \cite{qiu04} found that a maximum acceleration occurs at the time of maximum magnetic reconnection rate derived from the ribbon expansion, and claimed that the flux rope motion is affected by the magnetic field reconnection. As a support to this view, we have a couple of radio signatures. The DPS in the 0.8--1.4~GHz range during 16:37:15--16:41:00~UT suggests the ejection of the plasmoids \citep{karlicky05}, and the decametric type III radio burst immediately preceding a decametric type II burst implies that the flux rope blew the envelope field in the upper corona. The type III burst could be generated when the local reconnection started between the rising flux rope and surrounding magnetic field (Fig.~\ref{fig.9}{\it b}), in a similar picture presented in the emerging flux model of flares \citep{hey77} while possibly in a larger spatial scale in the present situation. Subsequently, the rapidly rising flux rope generated the type II burst and finally escaped to become the CME.

\section{SUMMARY}
In this paper, we have presented a multiwavelength study of the 2005 May 13 eruption of a sigmoidal active region, that turned out to be a set of versatile observational evidence for the Moore's model. While the Moore's model was originally based on the morphological changes of SXR sigmoids, the present study found a variety of features observed in a wide range of wavelengths: SXRs, UV/EUV, HXRs, radio, H$\alpha$, and white-light. They allowed us to study the eruption process of this sigmoid in more detail than previous studies of sigmoids, which helped us to identify the most relevant triggering mechanism with less ambiguity. Our results and conclusions are summarized as follows.

\begin{enumerate}
\item	An important finding, in relation to the eruption onset mechanisms, is the earlier start of the flare brightening in the core of the active region than that of the rising motion of the flux rope. Together with other evidence from dynamics of filament (\S~3.2) and flare ribbons (\S~3.4), this clearly supports the scenario that the eruption is triggered by a runaway internal tether-cutting reconnection in the sigmoid core. This rejects the competing eruption mechanism that the flux rope rises due to a loss of equilibrium and it later induces the tether cutting below the flux rope. The other mechanism, the external tether-cutting mechanism \citepalias[see][]{moore06} is also rejected because no remote brightening was found in the vicinity of the flaring region.

\item	Although sigmoid has been well known in the SXR community as a special magnetic structure prone to magnetic eruptions, this is perhaps the first case of high resolution observation of a sigmoid at EUV wavelength. The preflare sigmoid in this event had been unambiguously observed at 1~MK TRACE 171~\AA\ Fe\,{\sc ix/x} channel, 1.5~MK EIT 195~\AA\ Fe\,{\sc xii} channel, and 3.8~MK SXI polyimide thin filter position, thus confirming the structure visible in a wider temperature range than previously known \citep[$>$~1.5~MK, see][]{sterling00}. The high resolution preflare TRACE image clearly reveals not only the overall sigmoid shape, but also the highly sheared field in the core and the envelope coronal magnetic field. We schematically draw this structure in Figure~\ref{fig.9}{\it a}.

\item	The subsequent evolution of the sigmoidal active region up to the flare maximum as seen at EUV and SXRs is in good agreement with the Moore's model, and we quantitatively derive the magnetic reconnection rate to map out the slow runaway and explosive phase of the reconnection process. In specific, the initial flare brightening occurred in the middle of the sigmoid seems to signify the beginning of tether-cutting reconnection (Fig.~\ref{fig.9}{\it a}). Later expansion of the two elbows of the sigmoid indicates the ongoing slow runaway eruption. Finally, the envelope field is blown out by the explosive eruption of the flux rope near flare peak (Fig.~\ref{fig.9}{\it b}), which leads to the long-lived postflare reconnection and flare arcades.

\item	The filament was mostly not disrupted during this event. This implies that filament is neither the trigger nor essential for this kind of reconnection between sigmoidal loops. It is likely that the tether-cutting reconnection occurred above the strands of the sheared core field that held the filament and the resulting loops below the X-point actually protected the filament from disruption.

\item	The exceptional long-lived twin dimmings seen in this event, when placed under the context of the Moore's model, can be naturally explained by the eruption of the two magnetic elbows extending northeast and southwest of the active region (Fig.~\ref{fig.9}{\it b}).

\item	A special feature found in the RHESSI images is the morphological change of HXR emissions from the typical point-like compact sources (Fig.~\ref{fig.9}{\it a}, c.f. Fig.~\ref{fig.7}{\it a}--{\it b}) to the elongated ribbon-like source (Fig.~\ref{fig.9}{\it b}, c.f. Fig.~\ref{fig.7}{\it c}--{\it d}). We suggest that this footpoint-to-ribbon transformation of the HXR source is a natural outcome of the sigmoid-to-arcade evolution of the magnetic field configuration. 

\item	For this event we have a rich collection of radio emission features, including the type II precursors, DPS, and type II and III bursts. We interpret that the type II precursors are due to a low coronal shock driven by the rising flux rope seen at SXRs. The similar timing suggests that the decametric type II emission in this event occurred when the flux rope (CME) rapidly erupted near the flare maximum and interacted with the dense coronal streamer, which provides a condition in favor of the shock formation. Following \cite{karlicky05}, we associate the DPS in the 0.8--1.4~GHz with the ejection of the plasmoids around the flare maximum, and another broadband DPS in the 1.0--2.0~GHz range with the growing postflare arcades.
\end{enumerate}

\acknowledgments
The authors thank TRACE, SXI, BBSO, RHESSI, and SOHO teams for excellent data sets. We thank the referee for insightful comments that help to improve the paper. We are grateful to Stephen White and Henry Aurass for providing the radio data and having helpful discussions. CL is grateful to Steven Hill for help on SXI data processing, and to Guillermo A. Stenborg and Karl Battams for help on LASCO data processing. CL is indebted to Brian R. Dennis and A. Kimberley Tolbert for precious help on RHESSI software. This work is supported by NSF/SHINE grant ATM 05-48952. JL was supported by NSF grant AST 06-07544 and NASA grant NNG0-6GE76G. VY was supported by NSF grant ATM 05-36921 and NASA grant NNG0-4GJ51G. KSC has been supported by the MOST funds (M1-0104-00-0059 and M1-0407-00-0001) of the Korean government.

\end{document}